\documentclass[final,onecolumn,twoside,a4paper,10pt]{article}

\makeatletter
\def\section{\@startsection{section}{1}{\z@}%
    {-21dd plus-4pt minus-4pt}{10.5dd plus 4pt
     minus4pt}{\large\sffamily\bfseries}}
\def\subsection{\@startsection{subsection}{2}{\z@}%
    {-21dd plus-4pt minus-4pt}{10.5dd plus 4pt
     minus4pt}{\normalsize\sffamily\bfseries}}
\def\subsubsection{\@startsection{subsubsection}{3}{\z@}%
    {-21dd plus-4pt minus-4pt}{10.5dd plus 4pt
     minus4pt}{\normalsize\sffamily}}
\def\paragraph{\@startsection{subsubsection}{4}{\z@}%
    {-21dd plus-4pt minus-4pt}{-1pt}{\normalsize\sffamily\bfseries}}
\def\authorfont{\rmfamily}

\@settopoint\textwidth
\def\@listi{\leftmargin\leftmargini
            \parsep \z@
            \topsep 6\p@ \@plus2\p@ \@minus4\p@
            \itemsep\parsep}
\let\@listI\@listi
\@listi
\def\@listii {\leftmargin\leftmarginii
              \labelwidth\leftmarginii
              \advance\labelwidth-\labelsep
              \topsep    \z@
              \parsep    \topsep
              \itemsep   \parsep}
\def\@listiii{\leftmargin\leftmarginiii
              \labelwidth\leftmarginiii
              \advance\labelwidth-\labelsep
              \topsep    \z@
              \parsep    \topsep
              \itemsep   \parsep}
\def\@listiv {\leftmargin\leftmarginiv
              \labelwidth\leftmarginiv
              \advance\labelwidth-\labelsep}
\def\@listv  {\leftmargin\leftmarginv
              \labelwidth\leftmarginv
              \advance\labelwidth-\labelsep}
\def\@listvi {\leftmargin\leftmarginvi
              \labelwidth\leftmarginvi
              \advance\labelwidth-\labelsep}
\def\ps@headings{%
    \let\@oddfoot\@empty\let\@evenfoot\@empty
    \def\@evenhead{\small\rlap{\thepage}\hfil\leftmark\unskip}%
    \def\@oddhead{\small\rightmark\hfil\llap{\thepage}}%
    \let\@mkboth\@gobbletwo
    \let\sectionmark\@gobble
    \let\subsectionmark\@gobble
    }
\def\setitemindent#1{\settowidth{\labelwidth}{#1}%
        \leftmargini\labelwidth
        \advance\leftmargini\labelsep
   \def\@listi{\leftmargin\leftmargini
        \labelwidth\leftmargini\advance\labelwidth by -\labelsep
        \parsep=\parskip
        \topsep=\medskipamount
        \itemsep=\parskip \advance\itemsep by -\parsep}}
\def\setitemitemindent#1{\settowidth{\labelwidth}{#1}%
        \leftmarginii\labelwidth
        \advance\leftmarginii\labelsep
\def\@listii{\leftmargin\leftmarginii
        \labelwidth\leftmarginii\advance\labelwidth by -\labelsep
        \parsep=\parskip
        \topsep=\z@
        \itemsep=\parskip \advance\itemsep by -\parsep}}
\def\descriptionlabel#1{\hspace\labelsep #1\hfil}
\def\description{\@ifnextchar[{\@describe}{\list{}{\labelwidth\z@
          \itemindent-\leftmargin \let\makelabel\descriptionlabel}}}

\def\describelabel#1{#1\hfil}
\def\@describe[#1]{\relax\ifnum\@listdepth=0
\setitemindent{#1}\else\ifnum\@listdepth=1
\setitemitemindent{#1}\fi\fi
\list{--}{\let\makelabel\describelabel}}
\leftmargin  \leftmargini
\@beginparpenalty -\@lowpenalty
\@endparpenalty   -\@lowpenalty
\@itempenalty     -\@lowpenalty
\renewcommand\theenumi{\@arabic\c@enumi}
\renewcommand\theenumii{\@alph\c@enumii}
\renewcommand\theenumiii{\@roman\c@enumiii}
\renewcommand\theenumiv{\@Alph\c@enumiv}

\renewcommand\p@enumii{\theenumi}
\renewcommand\p@enumiii{\theenumi(\theenumii)}
\renewcommand\p@enumiv{\p@enumiii\theenumiii}

\renewcommand\labelitemiii{$\m@th\bullet$}
\renewcommand\labelitemiv{$\m@th\cdot$}

\def\abstract{\topsep=0pt\partopsep=0pt\parsep=0pt\itemsep=0pt\relax
\trivlist\item[\hskip\labelsep
{\bfseries\abstractname}]\if!\abstractname!\hskip-\labelsep\fi}
\if@twocolumn

\else
   
\fi

\makeatother

\usepackage{multicol}

\usepackage{listings}

\begin{document}

\def\abstracttext{
This article describes a methodology for fitting experimental data to the discrete 
power-law distribution and provides the results of a detailed simulation exercise 
used to calculate accurate cutoff values used to assess the fit to a power-law 
distribution when using the maximum likelihood estimation for the exponent 
of the distribution.  
Using massively parallel programming computing, we were able to accelerate 
by a factor of 60  the computational time required for these calculations 
across a range of parameters and construct a series of detailed tables containing 
the test values to be  used in a Kolmogorov-Smirnov goodness-of-fit test, allowing 
for an accurate assessment of the power-law fit from empirical data. 
}

\begin{multicols}{2}[
{
\normalfont
 {\LARGE \sffamily\bfseries
  \pretolerance=10000
  \rightskip=0pt plus 4cm
  \noindent\ignorespaces
{Using GPU Simulation to Accurately Fit to the \\Power-Law Distribution}}
\vskip 11.24pt\relax
 \authorfont
 \lineskip .5em
 \rightskip=0pt plus 2cm
 \noindent\ignorespaces  \Large
{Efstratios Rappos and Stephan Robert\footnote{{\it Contact information: }Haute Ecole d'Ing{\'e}nierie et de Gestion du Canton de Vaud, 
HEIG-VD, Route de Cheseaux 1, CH-1401 Yverdon-les-Bains, 
Switzerland; Efstratios.Rappos@heig-vd.ch, Stephan.Robert@heig-vd.ch}}
\vskip7.23pt
 \rightskip=0pt\relax
 \small\rm
\noindent
Institut des technologies de l'information et de la communication\\
{Haute Ecole d'Ing{\'e}nierie et de Gestion du Canton de Vaud (HEIG-VD)}\\
Haute Ecole Sp{\'e}cialis{\'e}e de Suisse occidentale
\\[\baselineskip]
Institute for Information and Communication Technologies\\
School of Business and Engineering Vaud  (HEIG-VD)\\
University of Applied Sciences of Western Switzerland
 \vskip 12.85pt
 \leftskip=1.5cm\rightskip=\leftskip
 \noindent
 \vskip 12.85pt
 {\topsep=0pt\partopsep=0pt\parsep=0pt\itemsep=0pt\relax
  \trivlist
\item[\hskip\labelsep{\sffamily\bfseries Abstract}] %
  \leftskip=1.5cm\rightskip=\leftskip
  \abstracttext
\\
\item[\hskip\labelsep{\sffamily\bfseries Keywords:}] Zipf distribution; power law; Kolmogorov-Smirnov test;  parallel computing simulation;
graphics processing unit programming \\
\item[\hskip\labelsep{\sffamily\bfseries PACS:\enspace}] 02.50.Ng--Distribution theory and Monte Carlo studies; 05.10.Ln--Monte Carlo methods; 89.75.-k--Complex systems\\
\item[\hskip\labelsep{\sffamily\bfseries MSC 2010:}]  62F03, 68W10, 62P10, 62P30, 62P35, 62Q05
\item \hfill {May 2013}
\endtrivlist
}
 \vskip22.47pt
}
]

\section{Introduction}

Power-law distributions and their extensions characterize many physical, biological and social phenomena \cite{clauset,newman,zornig,lee10b,su07,yeh02} but
the process of accurately fitting a power-law distribution to empirical data is not straightforward, and
in some cases very imprecise methods are known to be used, namely `estimating' the power-law 
exponent and fit via linear regression on a log-log plot \cite{clauset}.

A popular method to fit a power-law is by calculating the maximum likelihood estimator (MLE) for the 
distribution exponent and then using the Kolmogorov-Smirnov (KS) test to assess the goodness-of-fit 
by comparing against simulation-derived cutoff values. The practicalities of this
approach are described in \cite{clauset} and \cite{goldstein}. 

To produce these cutoff values, a large number of statistical simulations needs to be run. 
However, generic tables cannot always be used accurately, as the cutoff values depend on 
the sample size and the estimated value of the exponent of the data. 

Producing such tables for the power-law is computationally challenging. 
The most complete set of tables to date was produced by \cite{goldstein}; 
however, presumably due 
to limitations of the computer technology of the time, aggregate values 
were obtained across a range of values for the estimated exponent.  
We extend this work by providing the calculated cutoff tables for a variety of sample sizes and
values for the exponent, a task that would require over 2.5 years of computational time on 
a typical PC.  We also describe the methodology and provide computer code which enables
researchers to calculate the corresponding tables for values of the exponent other than the ones
we considered. 

Recent technological developments in the field of Graphics Processing Units (GPU), have 
resulted in consumer-level graphical cards being able to assist with computationally intensive 
tasks, because their massively parallel design can outperform traditional CPU algorithms. 
The use of graphics cards to improve the computational power for simulation methods has 
been studied in many areas such as Monte Carlo techniques \cite{lee10} and Bayesian estimation \cite{such10}. 
 
We demonstrate the use of GPU algorithms for the estimation of the KS cutoff values for 
assessing the goodness-of-fit of power-law data. The use of parallel methods allows 
much larger simulations to be produced in a shorter time, producing more accurate results and higher precision.

Furthermore, we consider the case of the truncated power-law distribution where there is an upper 
limit to the distribution values. This variation allows for cases where the exponent $\gamma<1$ to be fitted, as is the case in some phenomena such as the world-wide-web \cite{breslau}.

We consider two versions of the discrete power-low distribution, known as the Zipf distribution,
described by:
\begin{equation}
p(k) = \frac{k^{-\gamma}}{\zeta(\gamma)}  
\end{equation}
where
\begin{itemize}
\item $k$ is a positive integer 1, 2, 3, \ldots;
\item $p(k)$ is the probability of observing the value $k$;
\item $\gamma >1 $ is the power-law exponent;
\item $\zeta(\gamma)$ is an appropriate scaling factor.  
\end{itemize}

In the traditional version of the power-law the value of the integer $k$ is unbounded ($k\ge1$) and in 
that case the scaling factor is the Riemann zeta function $\zeta(\gamma) = \sum_{k=1}^\infty k^{-\gamma}$ and 
for convergence we must have $\gamma>1$. 

If we assume that the range of values for $k$ is finite i.e., $k=1, 2, \ldots, K$, then in this 
truncated Zipf distribution the scaling factor is $\zeta(\gamma) = \sum_{k=1}^K k^{-\gamma}$ and 
we only require the exponent to be $\gamma>0$ for convergence.

\section{Estimating the power-law exponent from the data}   \label{section-mle}

The maximum likelihood estimator for the power-law parameter is described in \cite{goldstein}
and applies to both variations of the Zipf distribution.
If the observed dataset consists of $N$ observations $x_1, x_2, \ldots, x_N$,  
the best estimate for $\gamma$ is the value that satisfies the equation
\begin{equation}
\frac{\zeta'(\gamma)}{\zeta(\gamma)} = - \frac{1}{N} \sum_{i=1}^N \log(x_i) \label{eq-mle}
\end{equation}
where $\zeta(\gamma)$ is either the scaling factor described in the previous section.
The above differential equation can easily be solved for $\gamma$ using the standard Newton-Raphson method.

\section{A KS goodness-of-fit test for power-law distributions}

The Kolmogorov-Smirnov test is a traditional statistical test for goodness-of-fit, relying 
on calculating the statistic
\begin{equation}
K = \sup_x |F^*(x) - S(x)|  \label{eq-ks}
\end{equation}
where $F^*$ is the hypothesized cumulative distribution function and $S$ is the empirical cumulative distribution based on the sample data, which is then compared with specific cutoff values. There are alternative approaches, such as the general Khmaladze transformation \cite{khm1,khm2}, but are outside 
of the scope of this article. 
The standard tables of cutoff values for the KS test cannot be directly used when the model parameters (the $\gamma$ in our case) have been estimated from the data, 
and bespoke tables have to be created using Monte-Carlo simulation.  
Moreover, the tables to be used also depend on the estimated value of $\gamma$ and the sample size. 

Cutoff values provided in \cite{goldstein} were obtained by simulating 10,000 Zipf distributions 
with a  random exponent $\gamma=1.5$ to $4.0$, for 14 logarithmically-spaced choices of the sample size.
Whilst this method produces reasonable results, we cannot ignore the fact that the KS cutoff values 
depend on the calculated value of $\gamma$ and therefore average values do not work well 
for cases where the power-law fit is marginal. 

We extend the results by providing the corresponding test values, simulating 50,000 Zipf distributions
for 15 similar choices of sample size, and in each case for 12 possible values of $\gamma$. 
In addition, we consider the case of the truncated distribution where observations are bounded at $K=20, 50, 100, 500$ and $1000$.  
We repeat each experiment 10 times for each case and tabulate the average value obtained in each case. 
In total, this results to a total of over 10,000 separate simulations compared to the 14 used in the 
above-mentioned research, each one containing five times the number of points. 

\section{A CUDA algorithm for the calculation of the KS test values}

To achieve this level of experimentation, the simulations were performed in a parallel computing environment consisting of two GTX590 graphics processing units (GPU) on a PC using 
the CUDA/C programming language. This approach carries out the calculations in a high-end 
computer graphics card rather than in the CPU and the inherent parallel architecture of the GPU
makes it well suited for simulation experimentation, allowing for a 60 times faster program execution 
speed compared to CPU calculations. Indeed, we were able to produce these simulation 
results in just over 373 hours of computational time; using traditional CPU programming this would have taken 2.5 years.   

The algorithm, available as a supplementary material to this article, separates the simulations 
into 782 blocks of 64 simulations (threads) each. The last 48 simulations are discarded to give the 
required 50,000 simulations. The program is repeated for the different values of $N$, $K$ and $\gamma$.
Care is taken in the code to ensure an efficient execution, for example, the natural logarithms of 
the first $K$ integers are pre-computed and stored in an array: this speeds up considerably the calculation since the terms $k^{-\gamma}$, which appear in $\zeta(\gamma)$ and its derivatives, 
can be calculated as $e^{-\gamma \ln k}$. Care should also be taken, as explained in the attached code,
to adjust a compiler parameter when running the code in order to ensure all calculations 
are carried out in double-precision rather than single-precision by default and  
avoid numerical underflow in the calculations. 

Table~\ref{table-1} presents the test values to use for the pure Zipf distribution (which corresponds 
to a truncated Zipf distribution with $K=\infty$) for various choices of the estimated value 
of the exponent $\gamma$. 
Tables~\ref{table-2} to~\ref{table-last} present the corresponding tables for the truncated power-law
distribution  with $K=20$, $50$, $100$, $500$ and $1000$ respectively.

These refinements extend the accuracy of the implementation. 
We note the variation in the cutoff values of Table~\ref{table-1} depending on
the exponent $\gamma$: for example, the 90\% cutoff value for a sample size of 1,000 ranges 
from 0.0056 when $\gamma=4$ to 0.0569 when $\gamma=1.25$, a difference of a factor of 10. 
In contrast, the corresponding figure in \cite{goldstein}, calculated for an `average' exponent 
is reported to be 0.0186. 
This demonstrates the importance of using cutoff tables that are particular not only to the specific 
sample size but also the value of the exponent $\gamma$. 

In practice, the value of $\gamma$ calculated from the data will probably not be an exact
match with any of the tabulated values. 
Ideally, to achieve the best level of accuracy, a  meticulous researcher would have to create 
a bespoke table containing the cutoff values that correspond to the exact value of $\gamma$ 
as calculated from the sample. 
Nevertheless, our tables provide a useful approximation for cases where this level of precision is not
required, and a simple gauge of how good the power-law fit is required. 
In any case, marginal cases aside, using these tables with a close approximate value for $\gamma$ 
can be a lot more precise than log-log plots or the Pearson's test.

Finally, it is worth noting that the tables presented apply only when the exponent $\gamma$ has been calculated using the MLE method 
described in Section~\ref{section-mle} and would not be relevant if a different method was used instead.

The way to use these tables in practice is described in  \cite{clauset} and \cite{goldstein}. 
Assuming one has a set of discrete observations and wishes to test if they follow the Zipf distribution, 
they would first calculate the maximum likelihood estimator for the exponent $\gamma$ using 
(\ref{eq-mle}). Then, they would calculate the test statistic (\ref{eq-ks}) by determining the maximum
deviation of the empirical cumulative distribution function against the theoretical Zipf one. 

This test statistic will then be compared with the cutoff value in the tables that corresponds to the 
values of $N$, $K$ and estimated $\gamma$ of the observed dataset. If the test value is less than 
the tabulated value, there is insufficient evidence to reject the hypothesis that the data follow a Zipf 
distribution, at the required level of significance. As mentioned earlier, for maximum accuracy a bespoke
cutoff value would ideally need to be calculated matching exactly the values of $N$, $K$, $\gamma$ 
of the sample. This can be achieved using the accompanying code.

\section{Conclusions}

We presented the results of a detailed simulation to calculate the cutoff values of the 
Kolmogorov-Smirnov test when used to assess the fit of empirical data 
to the discrete Zipf or power-law distribution. We carry out a much larger set of 
simulations that the state-of-the art and further extend previous research by 
breaking down the cutoff tables according to the estimated 
value of the Zipf exponent and further consider two versions of the Zipf distribution. 

This level of complexity was only 
possible using Graphical Processing Unit (GPU) algorithms to massively parallelize the simulations. 
In doing so, we produced a 60-fold faster simulation algorithm compared with traditional programming 
techniques, which demonstrates
the huge potential value of GPU techniques in improving the performance of statistical simulations 
and other complex algorithms. The provided computer code is also of benefit to any researcher who
needs, for more accuracy, to create their own Kolmogorov-Smirnov cutoff value which is specific to 
the sample size and estimated exponent of their datasets.

\section{Supplementary Materials}

\begin{description}
\item[CUDA/C code:] The annex contains the CUDA program that can be used to replicate
the results presented in this article. The instructions for compilation and use are included in the code.
\end{description}

\bibliographystyle{plain}
\bibliography{bib-rappos-robert}

\columnbreak

\end{multicols}

\begin{table*}																
\caption{KS test statistic for the pure power-law distribution}\label{table-1}		
\tiny																
\begin{tabular}{lrrrr|rrrr|rrrr|rrrr}
\hline\noalign{\smallskip}
& \multicolumn{4}{c|}{$\gamma=1.25$, Quantiles} & \multicolumn{4}{c|}{$\gamma=1.5$, Quantiles} & \multicolumn{4}{c|}{$\gamma=1.75$, Quantiles} & \multicolumn{4}{c}{$\gamma=2.0$, Quantiles} \\
$N$ & 0.9 & 0.95 & 0.99 & 0.999 & 0.9 & 0.95 & 0.99 & 0.999 & 0.9 & 0.95 & 0.99 & 0.999 & 0.9 & 0.95 & 0.99 & 0.999 \\	
\noalign{\smallskip}\hline\noalign{\smallskip}
10&	.2792&	.3092&	.3668&	.4315&	.2410&	.2710&	.3308&	.3994&	.2115&	.2411&	.3027&	.3761&	.1835&	.2116&	.2739&	.3515\\
20&	.2043&	.2262&	.2692&	.3198&	.1702&	.1913&	.2342&	.2865&	.1489&	.1694&	.2127&	.2656&	.1289&	.1484&	.1897&	.2426\\
30&	.1716&	.1896&	.2251&	.2668&	.1392&	.1564&	.1916&	.2346&	.1217&	.1383&	.1739&	.2173&	.1054&	.1207&	.1541&	.1960\\
40&	.1522&	.1678&	.1993&	.2353&	.1207&	.1354&	.1665&	.2048&	.1054&	.1195&	.1504&	.1893&	.0911&	.1045&	.1331&	.1697\\
50&	.1391&	.1532&	.1808&	.2136&	.1080&	.1211&	.1491&	.1834&	.0943&	.1071&	.1343&	.1688&	.0815&	.0934&	.1189&	.1526\\
100&	.1073&	.1174&	.1373&	.1610&	.0766&	.0860&	.1058&	.1307&	.0666&	.0756&	.0949&	.1192&	.0576&	.0658&	.0836&	.1062\\
500&	.0662&	.0708&	.0799&	.0907&	.0352&	.0396&	.0485&	.0597&	.0298&	.0338&	.0424&	.0533&	.0258&	.0294&	.0373&	.0468\\
1000&	.0569&	.0602&	.0667&	.0742&	.0257&	.0288&	.0353&	.0433&	.0211&	.0239&	.0299&	.0376&	.0182&	.0208&	.0264&	.0334\\
2000&	.0505&	.0529&	.0575&	.0629&	.0192&	.0215&	.0261&	.0317&	.0149&	.0169&	.0212&	.0266&	.0129&	.0147&	.0187&	.0236\\
3000&	.0478&	.0497&	.0535&	.0579&	.0164&	.0183&	.0220&	.0266&	.0122&	.0138&	.0173&	.0217&	.0105&	.0120&	.0152&	.0192\\
4000&	.0461&	.0478&	.0511&	.0548&	.0148&	.0164&	.0196&	.0237&	.0106&	.0120&	.0150&	.0189&	.0091&	.0104&	.0131&	.0167\\
5000&	.0450&	.0466&	.0495&	.0529&	.0137&	.0151&	.0181&	.0217&	.0095&	.0107&	.0135&	.0168&	.0081&	.0093&	.0118&	.0149\\
10000&	.0424&	.0435&	.0456&	.0479&	.0109&	.0119&	.0140&	.0165&	.0067&	.0076&	.0095&	.0120&	.0058&	.0066&	.0083&	.0105\\
20000&	.0406&	.0413&	.0428&	.0445&	.0090&	.0098&	.0112&	.0130&	.0048&	.0054&	.0068&	.0085&	.0041&	.0047&	.0059&	.0075\\
50000&	.0390&	.0395&	.0405&	.0415&	.0074&	.0078&	.0087&	.0098&	.0031&	.0035&	.0044&	.0055&	.0026&	.0029&	.0037&	.0047\\
\noalign{\smallskip}\hline\noalign{\smallskip}		
& \multicolumn{4}{c|}{$\gamma=2.5$, Quantiles} & \multicolumn{4}{c|}{$\gamma=3.0$, Quantiles} & \multicolumn{4}{c|}{$\gamma=3.5$, Quantiles} & \multicolumn{4}{c}{$\gamma=4.0$, Quantiles} \\	
$N$ & 0.9 & 0.95 & 0.99 & 0.999 & 0.9 & 0.95 & 0.99 & 0.999 & 0.9 & 0.95 & 0.99 & 0.999 & 0.9 & 0.95 & 0.99 & 0.999 \\																
\noalign{\smallskip}\hline\noalign{\smallskip}																
10&	.1375&	.1591&	.2141&	.2876&	.1005&	.1281&	.1695&	.2351&	.0819&	.0992&	.1365&	.1921&	.0819&	.0819&	.1092&	.1530\\
20&	.0963&	.1117&	.1472&	.1936&	.0724&	.0849&	.1123&	.1528&	.0534&	.0643&	.0879&	.1245&	.0440&	.0533&	.0735&	.0995\\
30&	.0781&	.0903&	.1175&	.1540&	.0573&	.0678&	.0897&	.1188&	.0430&	.0532&	.0699&	.0950&	.0311&	.0409&	.0550&	.0763\\
40&	.0674&	.0782&	.1012&	.1314&	.0501&	.0584&	.0771&	.1014&	.0372&	.0444&	.0593&	.0800&	.0281&	.0351&	.0463&	.0639\\
50&	.0603&	.0697&	.0901&	.1173&	.0448&	.0521&	.0679&	.0898&	.0330&	.0391&	.0526&	.0704&	.0264&	.0306&	.0408&	.0552\\
100&	.0427&	.0491&	.0630&	.0815&	.0315&	.0365&	.0471&	.0608&	.0235&	.0273&	.0357&	.0467&	.0178&	.0207&	.0277&	.0369\\
500&	.0190&	.0219&	.0280&	.0353&	.0141&	.0162&	.0207&	.0261&	.0105&	.0121&	.0155&	.0196&	.0079&	.0092&	.0118&	.0151\\
1000&	.0135&	.0155&	.0198&	.0251&	.0100&	.0115&	.0146&	.0185&	.0074&	.0086&	.0109&	.0137&	.0056&	.0064&	.0083&	.0105\\
2000&	.0095&	.0109&	.0140&	.0177&	.0070&	.0081&	.0103&	.0130&	.0052&	.0060&	.0077&	.0097&	.0039&	.0046&	.0058&	.0073\\
3000&	.0078&	.0089&	.0114&	.0144&	.0057&	.0066&	.0084&	.0106&	.0043&	.0049&	.0063&	.0079&	.0032&	.0037&	.0047&	.0059\\
4000&	.0067&	.0077&	.0098&	.0125&	.0050&	.0057&	.0073&	.0092&	.0037&	.0043&	.0054&	.0068&	.0028&	.0032&	.0041&	.0052\\
5000&	.0060&	.0069&	.0088&	.0112&	.0045&	.0051&	.0065&	.0082&	.0033&	.0038&	.0048&	.0061&	.0025&	.0029&	.0037&	.0046\\
10000&	.0043&	.0049&	.0062&	.0079&	.0031&	.0036&	.0046&	.0058&	.0023&	.0027&	.0034&	.0043&	.0018&	.0020&	.0026&	.0032\\
20000&	.0030&	.0035&	.0044&	.0056&	.0022&	.0026&	.0033&	.0041&	.0017&	.0019&	.0024&	.0030&	.0012&	.0014&	.0018&	.0023\\
50000&	.0019&	.0022&	.0028&	.0035&	.0014&	.0016&	.0021&	.0026&	.0010&	.0012&	.0015&	.0019&	.0008&	.0009&	.0012&	.0015\\
\noalign{\smallskip}\hline																
\end{tabular}																
\end{table*}


\begin{table*}																
\caption{KS test statistic for the truncated power-law distribution with $K=20$ }\label{table-2}		
\tiny																
\begin{tabular}{lrrrr|rrrr|rrrr|rrrr}															
\hline\noalign{\smallskip}																
& \multicolumn{4}{c|}{$K=20$: $\gamma=0.25$, Quantiles} & \multicolumn{4}{c|}{$\gamma=0.5$, Quantiles} & \multicolumn{4}{c|}{$\gamma=0.75$, Quantiles} & \multicolumn{4}{c}{$\gamma=1.0$, Quantiles} \\	
$N$ & 0.9 & 0.95 & 0.99 & 0.999 & 0.9 & 0.95 & 0.99 & 0.999 & 0.9 & 0.95 & 0.99 & 0.999 & 0.9 & 0.95 & 0.99 & 0.999 \\																
\noalign{\smallskip}\hline\noalign{\smallskip}														
10&	.2486&	.2751&	.3286&	.3915&	.2387&	.2640&	.3159&	.3770&	.2266&	.2503&	.2990&	.3595&	.2128&	.2353&	.2812&	.3387\\
20&	.1752&	.1943&	.2336&	.2796&	.1684&	.1865&	.2239&	.2684&	.1599&	.1767&	.2114&	.2530&	.1504&	.1662&	.1983&	.2380\\
30&	.1436&	.1594&	.1914&	.2310&	.1376&	.1525&	.1828&	.2197&	.1303&	.1442&	.1722&	.2062&	.1224&	.1354&	.1615&	.1932\\
40&	.1245&	.1381&	.1664&	.2006&	.1195&	.1323&	.1587&	.1915&	.1131&	.1252&	.1495&	.1796&	.1061&	.1174&	.1400&	.1682\\
50&	.1114&	.1237&	.1490&	.1792&	.1067&	.1183&	.1425&	.1702&	.1012&	.1119&	.1339&	.1602&	.0949&	.1049&	.1252&	.1504\\
100&	.0788&	.0876&	.1054&	.1267&	.0755&	.0838&	.1007&	.1212&	.0716&	.0792&	.0947&	.1137&	.0671&	.0742&	.0886&	.1059\\
500&	.0352&	.0392&	.0471&	.0569&	.0338&	.0375&	.0451&	.0543&	.0320&	.0354&	.0424&	.0508&	.0300&	.0332&	.0396&	.0474\\
1000&	.0249&	.0277&	.0333&	.0403&	.0239&	.0265&	.0318&	.0387&	.0226&	.0250&	.0299&	.0361&	.0212&	.0235&	.0280&	.0334\\
2000&	.0176&	.0196&	.0236&	.0285&	.0169&	.0187&	.0225&	.0271&	.0160&	.0177&	.0212&	.0254&	.0150&	.0166&	.0198&	.0236\\
3000&	.0144&	.0160&	.0192&	.0233&	.0138&	.0153&	.0184&	.0221&	.0130&	.0144&	.0173&	.0208&	.0122&	.0135&	.0161&	.0193\\
4000&	.0125&	.0139&	.0167&	.0202&	.0119&	.0133&	.0159&	.0192&	.0113&	.0125&	.0150&	.0180&	.0106&	.0117&	.0140&	.0167\\
5000&	.0111&	.0124&	.0149&	.0181&	.0107&	.0119&	.0142&	.0171&	.0101&	.0112&	.0134&	.0160&	.0095&	.0105&	.0125&	.0150\\
10000&	.0079&	.0088&	.0106&	.0128&	.0076&	.0084&	.0101&	.0121&	.0072&	.0079&	.0095&	.0114&	.0067&	.0074&	.0089&	.0106\\
20000&	.0056&	.0062&	.0075&	.0091&	.0053&	.0059&	.0071&	.0086&	.0051&	.0056&	.0067&	.0081&	.0047&	.0052&	.0063&	.0075\\
50000&	.0035&	.0039&	.0047&	.0057&	.0034&	.0037&	.0045&	.0054&	.0032&	.0035&	.0042&	.0051&	.0030&	.0033&	.0039&	.0047\\
\noalign{\smallskip}																
\hline\noalign{\smallskip}																
& \multicolumn{4}{c|}{$K=20$: $\gamma=1.25$, Quantiles} & \multicolumn{4}{c|}{$\gamma=1.5$, Quantiles} & \multicolumn{4}{c|}{$\gamma=1.75$, Quantiles} & \multicolumn{4}{c}{$\gamma=2.0$, Quantiles} \\
$N$ & 0.9 & 0.95 & 0.99 & 0.999 & 0.9 & 0.95 & 0.99 & 0.999 & 0.9 & 0.95 & 0.99 & 0.999 & 0.9 & 0.95 & 0.99 & 0.999 \\																
\noalign{\smallskip}\hline\noalign{\smallskip}														
10&	.1985&	.2205&	.2649&	.3222&	.1836&	.2053&	.2509&	.3115&	.1695&	.1901&	.2347&	.3001&	.1531&	.1727&	.2183&	.2869\\
20&	.1400&	.1554&	.1865&	.2263&	.1295&	.1444&	.1761&	.2173&	.1189&	.1336&	.1653&	.2070&	.1077&	.1226&	.1535&	.1962\\
30&	.1143&	.1267&	.1519&	.1833&	.1058&	.1179&	.1434&	.1769&	.0968&	.1089&	.1345&	.1688&	.0880&	.0998&	.1249&	.1586\\
40&	.0989&	.1096&	.1313&	.1587&	.0915&	.1020&	.1241&	.1524&	.0838&	.0943&	.1164&	.1467&	.0761&	.0863&	.1081&	.1377\\
50&	.0884&	.0981&	.1175&	.1423&	.0817&	.0912&	.1109&	.1363&	.0749&	.0843&	.1040&	.1310&	.0681&	.0771&	.0962&	.1228\\
100&	.0624&	.0693&	.0833&	.1005&	.0577&	.0643&	.0783&	.0963&	.0530&	.0596&	.0735&	.0916&	.0480&	.0544&	.0680&	.0855\\
500&	.0279&	.0310&	.0371&	.0449&	.0258&	.0288&	.0350&	.0432&	.0236&	.0265&	.0327&	.0409&	.0215&	.0243&	.0303&	.0379\\
1000&	.0197&	.0219&	.0263&	.0317&	.0182&	.0203&	.0248&	.0305&	.0167&	.0188&	.0232&	.0291&	.0152&	.0172&	.0215&	.0271\\
2000&	.0140&	.0155&	.0186&	.0224&	.0129&	.0144&	.0174&	.0214&	.0118&	.0133&	.0164&	.0204&	.0107&	.0122&	.0152&	.0190\\
3000&	.0114&	.0126&	.0151&	.0183&	.0105&	.0117&	.0143&	.0175&	.0096&	.0108&	.0134&	.0166&	.0088&	.0099&	.0124&	.0156\\
4000&	.0099&	.0109&	.0132&	.0159&	.0091&	.0102&	.0124&	.0153&	.0083&	.0094&	.0116&	.0145&	.0076&	.0086&	.0107&	.0135\\
5000&	.0088&	.0098&	.0118&	.0142&	.0082&	.0091&	.0111&	.0136&	.0075&	.0084&	.0104&	.0129&	.0068&	.0077&	.0096&	.0121\\
10000&	.0062&	.0069&	.0083&	.0100&	.0058&	.0064&	.0078&	.0096&	.0053&	.0059&	.0073&	.0091&	.0048&	.0054&	.0068&	.0085\\
20000&	.0044&	.0049&	.0059&	.0071&	.0041&	.0045&	.0055&	.0068&	.0037&	.0042&	.0052&	.0065&	.0034&	.0038&	.0048&	.0060\\
50000&	.0028&	.0031&	.0037&	.0045&	.0026&	.0029&	.0035&	.0043&	.0024&	.0027&	.0033&	.0041&	.0021&	.0024&	.0030&	.0038\\
\noalign{\smallskip}																
\hline\noalign{\smallskip}																
& \multicolumn{4}{c|}{$K=20$: $\gamma=2.5$, Quantiles} & \multicolumn{4}{c|}{$\gamma=3.0$, Quantiles} & \multicolumn{4}{c|}{$\gamma=3.5$, Quantiles} & \multicolumn{4}{c}{$\gamma=4.0$, Quantiles} \\		
$N$ & 0.9 & 0.95 & 0.99 & 0.999 & 0.9 & 0.95 & 0.99 & 0.999 & 0.9 & 0.95 & 0.99 & 0.999 & 0.9 & 0.95 & 0.99 & 0.999 \\																
\noalign{\smallskip}\hline\noalign{\smallskip}														
10&	.1240&	.1447&	.1867&	.2474&	.0956&	.1170&	.1519&	.2007&	.0821&	.0955&	.1310&	.1726&	.0821&	.0821&	.1074&	.1455\\
20&	.0871&	.1003&	.1290&	.1662&	.0695&	.0801&	.1048&	.1363&	.0524&	.0630&	.0852&	.1129&	.0440&	.0523&	.0722&	.0955\\
30&	.0706&	.0814&	.1030&	.1330&	.0547&	.0651&	.0839&	.1086&	.0421&	.0519&	.0669&	.0878&	.0310&	.0411&	.0539&	.0732\\
40&	.0612&	.0702&	.0891&	.1137&	.0476&	.0552&	.0721&	.0928&	.0363&	.0438&	.0567&	.0748&	.0280&	.0351&	.0456&	.0617\\
50&	.0547&	.0627&	.0797&	.1015&	.0427&	.0493&	.0635&	.0822&	.0326&	.0388&	.0506&	.0661&	.0263&	.0302&	.0403&	.0531\\
100&	.0386&	.0442&	.0559&	.0712&	.0301&	.0347&	.0443&	.0563&	.0231&	.0267&	.0346&	.0447&	.0178&	.0206&	.0272&	.0360\\
500&	.0172&	.0197&	.0248&	.0312&	.0134&	.0154&	.0195&	.0245&	.0103&	.0119&	.0151&	.0190&	.0078&	.0091&	.0117&	.0149\\
1000&	.0122&	.0140&	.0176&	.0222&	.0095&	.0109&	.0138&	.0173&	.0073&	.0084&	.0106&	.0134&	.0055&	.0064&	.0082&	.0104\\
2000&	.0086&	.0099&	.0124&	.0157&	.0067&	.0077&	.0098&	.0123&	.0051&	.0059&	.0075&	.0095&	.0039&	.0045&	.0058&	.0073\\
3000&	.0070&	.0080&	.0101&	.0128&	.0055&	.0063&	.0080&	.0100&	.0042&	.0048&	.0061&	.0077&	.0032&	.0037&	.0047&	.0059\\
4000&	.0061&	.0070&	.0088&	.0111&	.0047&	.0054&	.0069&	.0086&	.0036&	.0042&	.0053&	.0066&	.0028&	.0032&	.0041&	.0051\\
5000&	.0054&	.0062&	.0079&	.0099&	.0042&	.0049&	.0062&	.0078&	.0032&	.0037&	.0047&	.0059&	.0025&	.0029&	.0036&	.0046\\
10000&	.0039&	.0044&	.0056&	.0070&	.0030&	.0034&	.0044&	.0054&	.0023&	.0026&	.0033&	.0042&	.0017&	.0020&	.0026&	.0032\\
20000&	.0027&	.0031&	.0039&	.0049&	.0021&	.0024&	.0031&	.0039&	.0016&	.0019&	.0024&	.0030&	.0012&	.0014&	.0018&	.0023\\
50000&	.0017&	.0020&	.0025&	.0031&	.0013&	.0015&	.0020&	.0024&	.0010&	.0012&	.0015&	.0019&	.0008&	.0009&	.0011&	.0014\\
\noalign{\smallskip}\hline																
\end{tabular}																
\end{table*}


\begin{table*}																
\caption{KS test statistic for the truncated power-law distribution with $K=50$ }		
\tiny																
\begin{tabular}{lrrrr|rrrr|rrrr|rrrr}																
\hline\noalign{\smallskip}																
& \multicolumn{4}{c|}{$K=50$: $\gamma=0.25$, Quantiles} & \multicolumn{4}{c|}{$\gamma=0.5$, Quantiles} & \multicolumn{4}{c|}{$\gamma=0.75$, Quantiles} & \multicolumn{4}{c}{$\gamma=1.0$, Quantiles} \\
$N$ & 0.9 & 0.95 & 0.99 & 0.999 & 0.9 & 0.95 & 0.99 & 0.999 & 0.9 & 0.95 & 0.99 & 0.999 & 0.9 & 0.95 & 0.99 & 0.999 \\																
\noalign{\smallskip}\hline\noalign{\smallskip}														
10&	.2685&	.2960&	.3508&	.4146&	.2572&	.2833&	.3362&	.3985&	.2421&	.2660&	.3157&	.3755&	.2258&	.2479&	.2928&	.3504\\
20&	.1917&	.2113&	.2513&	.2986&	.1834&	.2021&	.2405&	.2853&	.1724&	.1895&	.2248&	.2665&	.1606&	.1763&	.2080&	.2475\\
30&	.1567&	.1730&	.2058&	.2451&	.1501&	.1653&	.1968&	.2349&	.1410&	.1550&	.1837&	.2193&	.1312&	.1441&	.1701&	.2024\\
40&	.1357&	.1498&	.1788&	.2136&	.1300&	.1434&	.1707&	.2042&	.1222&	.1344&	.1593&	.1902&	.1137&	.1248&	.1473&	.1758\\
50&	.1208&	.1336&	.1597&	.1906&	.1160&	.1280&	.1529&	.1819&	.1092&	.1201&	.1427&	.1698&	.1015&	.1116&	.1321&	.1568\\
100&	.0857&	.0948&	.1134&	.1355&	.0821&	.0907&	.1081&	.1292&	.0771&	.0849&	.1007&	.1203&	.0717&	.0788&	.0932&	.1107\\
500&	.0383&	.0424&	.0507&	.0607&	.0367&	.0405&	.0483&	.0578&	.0345&	.0380&	.0451&	.0538&	.0321&	.0352&	.0417&	.0494\\
1000&	.0271&	.0300&	.0358&	.0430&	.0259&	.0286&	.0341&	.0410&	.0244&	.0269&	.0318&	.0382&	.0227&	.0249&	.0294&	.0350\\
2000&	.0192&	.0212&	.0254&	.0304&	.0183&	.0203&	.0242&	.0289&	.0172&	.0190&	.0225&	.0269&	.0160&	.0176&	.0208&	.0248\\
3000&	.0156&	.0173&	.0207&	.0249&	.0150&	.0165&	.0197&	.0237&	.0141&	.0155&	.0184&	.0219&	.0131&	.0144&	.0170&	.0202\\
4000&	.0136&	.0150&	.0179&	.0216&	.0130&	.0143&	.0171&	.0205&	.0122&	.0134&	.0159&	.0190&	.0113&	.0125&	.0148&	.0175\\
5000&	.0121&	.0134&	.0161&	.0192&	.0116&	.0128&	.0153&	.0183&	.0109&	.0120&	.0143&	.0171&	.0101&	.0112&	.0132&	.0156\\
10000&	.0086&	.0095&	.0113&	.0136&	.0082&	.0091&	.0108&	.0129&	.0077&	.0085&	.0101&	.0120&	.0072&	.0079&	.0093&	.0110\\
20000&	.0061&	.0067&	.0080&	.0096&	.0058&	.0064&	.0077&	.0092&	.0055&	.0060&	.0071&	.0085&	.0051&	.0056&	.0066&	.0078\\
50000&	.0038&	.0042&	.0051&	.0061&	.0037&	.0041&	.0048&	.0058&	.0034&	.0038&	.0045&	.0054&	.0032&	.0035&	.0042&	.0050\\
\noalign{\smallskip}												
\hline\noalign{\smallskip}																
& \multicolumn{4}{c|}{$K=50$: $\gamma=1.25$, Quantiles} & \multicolumn{4}{c|}{$\gamma=1.5$, Quantiles} & \multicolumn{4}{c|}{$\gamma=1.75$, Quantiles} & \multicolumn{4}{c}{$\gamma=2.0$, Quantiles} \\
$N$ & 0.9 & 0.95 & 0.99 & 0.999 & 0.9 & 0.95 & 0.99 & 0.999 & 0.9 & 0.95 & 0.99 & 0.999 & 0.9 & 0.95 & 0.99 & 0.999 \\																
\noalign{\smallskip}\hline\noalign{\smallskip}														
10&	.2104&	.2323&	.2775&	.3335&	.1953&	.2180&	.2670&	.3307&	.1805&	.2035&	.2546&	.3234&	.1634&	.1863&	.2384&	.3105\\
20&	.1492&	.1646&	.1962&	.2373&	.1383&	.1541&	.1877&	.2342&	.1273&	.1438&	.1794&	.2252&	.1155&	.1318&	.1667&	.2139\\
30&	.1219&	.1345&	.1603&	.1931&	.1131&	.1261&	.1539&	.1902&	.1040&	.1174&	.1468&	.1845&	.0944&	.1075&	.1362&	.1727\\
40&	.1056&	.1165&	.1388&	.1672&	.0978&	.1091&	.1331&	.1652&	.0901&	.1017&	.1268&	.1610&	.0817&	.0930&	.1178&	.1512\\
50&	.0943&	.1041&	.1241&	.1497&	.0874&	.0976&	.1192&	.1472&	.0805&	.0909&	.1135&	.1435&	.0730&	.0832&	.1051&	.1348\\
100&	.0666&	.0737&	.0879&	.1061&	.0618&	.0690&	.0843&	.1046&	.0569&	.0643&	.0801&	.1014&	.0516&	.0587&	.0741&	.0941\\
500&	.0298&	.0329&	.0393&	.0475&	.0276&	.0308&	.0377&	.0470&	.0254&	.0287&	.0358&	.0452&	.0231&	.0263&	.0332&	.0417\\
1000&	.0211&	.0233&	.0278&	.0336&	.0195&	.0218&	.0267&	.0332&	.0180&	.0203&	.0254&	.0321&	.0163&	.0186&	.0234&	.0297\\
2000&	.0149&	.0164&	.0196&	.0238&	.0138&	.0154&	.0189&	.0235&	.0127&	.0143&	.0179&	.0225&	.0115&	.0131&	.0166&	.0209\\
3000&	.0122&	.0134&	.0160&	.0195&	.0113&	.0126&	.0154&	.0190&	.0104&	.0117&	.0146&	.0184&	.0094&	.0107&	.0135&	.0171\\
4000&	.0105&	.0116&	.0139&	.0169&	.0098&	.0109&	.0134&	.0166&	.0090&	.0101&	.0127&	.0159&	.0082&	.0093&	.0117&	.0148\\
5000&	.0094&	.0104&	.0124&	.0150&	.0087&	.0097&	.0119&	.0148&	.0080&	.0091&	.0113&	.0142&	.0073&	.0083&	.0105&	.0133\\
10000&	.0067&	.0074&	.0088&	.0106&	.0062&	.0069&	.0085&	.0104&	.0057&	.0064&	.0080&	.0101&	.0052&	.0059&	.0074&	.0094\\
20000&	.0047&	.0052&	.0062&	.0075&	.0044&	.0049&	.0060&	.0074&	.0040&	.0045&	.0057&	.0071&	.0036&	.0042&	.0052&	.0066\\
50000&	.0030&	.0033&	.0039&	.0047&	.0028&	.0031&	.0038&	.0047&	.0025&	.0029&	.0036&	.0045&	.0023&	.0026&	.0033&	.0042\\
\noalign{\smallskip}													
\hline\noalign{\smallskip}																
& \multicolumn{4}{c|}{$K=50$: $\gamma=2.5$, Quantiles} & \multicolumn{4}{c|}{$\gamma=3.0$, Quantiles} & \multicolumn{4}{c|}{$\gamma=3.5$, Quantiles} & \multicolumn{4}{c}{$\gamma=4.0$, Quantiles} \\		
$N$ & 0.9 & 0.95 & 0.99 & 0.999 & 0.9 & 0.95 & 0.99 & 0.999 & 0.9 & 0.95 & 0.99 & 0.999 & 0.9 & 0.95 & 0.99 & 0.999 \\																
\noalign{\smallskip}\hline\noalign{\smallskip}														
10&	.1307&	.1508&	.2037&	.2672&	.0993&	.1243&	.1612&	.2154&	.0819&	.0983&	.1350&	.1860&	.0819&	.0819&	.1085&	.1510\\
20&	.0920&	.1067&	.1373&	.1791&	.0715&	.0836&	.1096&	.1461&	.0532&	.0642&	.0872&	.1217&	.0440&	.0530&	.0732&	.0984\\
30&	.0747&	.0863&	.1107&	.1433&	.0566&	.0672&	.0881&	.1146&	.0428&	.0530&	.0693&	.0929&	.0311&	.0409&	.0550&	.0759\\
40&	.0646&	.0744&	.0955&	.1227&	.0494&	.0574&	.0750&	.0981&	.0371&	.0442&	.0587&	.0782&	.0281&	.0351&	.0463&	.0635\\
50&	.0578&	.0665&	.0852&	.1094&	.0441&	.0512&	.0666&	.0867&	.0329&	.0390&	.0521&	.0688&	.0264&	.0305&	.0407&	.0548\\
100&	.0408&	.0468&	.0598&	.0770&	.0311&	.0359&	.0461&	.0591&	.0234&	.0272&	.0354&	.0460&	.0178&	.0207&	.0276&	.0367\\
500&	.0182&	.0209&	.0266&	.0335&	.0139&	.0160&	.0203&	.0256&	.0104&	.0121&	.0154&	.0194&	.0079&	.0092&	.0118&	.0150\\
1000&	.0129&	.0148&	.0188&	.0238&	.0098&	.0113&	.0143&	.0181&	.0074&	.0085&	.0108&	.0136&	.0055&	.0064&	.0083&	.0104\\
2000&	.0091&	.0105&	.0133&	.0168&	.0069&	.0080&	.0101&	.0128&	.0052&	.0060&	.0077&	.0096&	.0039&	.0045&	.0058&	.0073\\
3000&	.0074&	.0085&	.0109&	.0137&	.0057&	.0065&	.0083&	.0104&	.0043&	.0049&	.0062&	.0078&	.0032&	.0037&	.0047&	.0059\\
4000&	.0064&	.0074&	.0094&	.0119&	.0049&	.0056&	.0072&	.0090&	.0037&	.0043&	.0054&	.0067&	.0028&	.0032&	.0041&	.0052\\
5000&	.0058&	.0066&	.0084&	.0107&	.0044&	.0050&	.0064&	.0081&	.0033&	.0038&	.0048&	.0060&	.0025&	.0029&	.0037&	.0046\\
10000&	.0041&	.0047&	.0059&	.0075&	.0031&	.0036&	.0045&	.0057&	.0023&	.0027&	.0034&	.0043&	.0018&	.0020&	.0026&	.0032\\
20000&	.0029&	.0033&	.0042&	.0053&	.0022&	.0025&	.0032&	.0040&	.0017&	.0019&	.0024&	.0030&	.0012&	.0014&	.0018&	.0023\\
50000&	.0018&	.0021&	.0027&	.0034&	.0014&	.0016&	.0020&	.0026&	.0010&	.0012&	.0015&	.0019&	.0008&	.0009&	.0012&	.0015\\
\noalign{\smallskip}\hline																
\end{tabular}
\end{table*}


\begin{table*}																
\caption{KS test statistic for the truncated power-law distribution with $K=100$ }	
\tiny																
\begin{tabular}{lrrrr|rrrr|rrrr|rrrr}
\hline\noalign{\smallskip}																
& \multicolumn{4}{c|}{$K=100$: $\gamma=0.25$, Quantiles} & \multicolumn{4}{c|}{$\gamma=0.5$, Quantiles} & \multicolumn{4}{c|}{$\gamma=0.75$, Quantiles} & \multicolumn{4}{c}{$\gamma=1.0$, Quantiles} \\	
$N$ & 0.9 & 0.95 & 0.99 & 0.999 & 0.9 & 0.95 & 0.99 & 0.999 & 0.9 & 0.95 & 0.99 & 0.999 & 0.9 & 0.95 & 0.99 & 0.999 \\																
\noalign{\smallskip}\hline\noalign{\smallskip}																
10&	.2772&	.3052&	.3602&	.4243&	.2657&	.2922&	.3460&	.4089&	.2490&	.2732&	.3235&	.3846&	.2313&	.2531&	.2978&	.3546\\
20&	.1990&	.2190&	.2596&	.3076&	.1905&	.2095&	.2483&	.2940&	.1781&	.1955&	.2312&	.2736&	.1649&	.1806&	.2124&	.2508\\
30&	.1631&	.1798&	.2132&	.2538&	.1562&	.1719&	.2037&	.2427&	.1462&	.1604&	.1896&	.2247&	.1351&	.1480&	.1737&	.2053\\
40&	.1416&	.1559&	.1855&	.2218&	.1355&	.1491&	.1772&	.2113&	.1268&	.1393&	.1644&	.1963&	.1172&	.1283&	.1506&	.1785\\
50&	.1265&	.1397&	.1662&	.1975&	.1213&	.1337&	.1588&	.1887&	.1135&	.1246&	.1474&	.1750&	.1049&	.1148&	.1349&	.1594\\
100&	.0893&	.0986&	.1174&	.1402&	.0857&	.0945&	.1122&	.1337&	.0802&	.0882&	.1043&	.1240&	.0742&	.0813&	.0956&	.1130\\
500&	.0400&	.0441&	.0526&	.0629&	.0383&	.0422&	.0502&	.0600&	.0358&	.0394&	.0466&	.0555&	.0331&	.0363&	.0427&	.0506\\
1000&	.0283&	.0312&	.0372&	.0444&	.0271&	.0299&	.0355&	.0425&	.0253&	.0278&	.0329&	.0393&	.0234&	.0256&	.0302&	.0355\\
2000&	.0200&	.0221&	.0263&	.0315&	.0192&	.0211&	.0251&	.0301&	.0179&	.0197&	.0233&	.0278&	.0166&	.0182&	.0213&	.0252\\
3000&	.0163&	.0180&	.0214&	.0257&	.0156&	.0172&	.0205&	.0245&	.0146&	.0161&	.0190&	.0226&	.0135&	.0148&	.0174&	.0205\\
4000&	.0141&	.0156&	.0186&	.0223&	.0135&	.0149&	.0178&	.0212&	.0127&	.0139&	.0165&	.0196&	.0117&	.0128&	.0151&	.0178\\
5000&	.0126&	.0140&	.0166&	.0199&	.0121&	.0134&	.0159&	.0190&	.0113&	.0125&	.0148&	.0175&	.0105&	.0115&	.0135&	.0159\\
10000&	.0090&	.0099&	.0118&	.0141&	.0086&	.0094&	.0112&	.0134&	.0080&	.0088&	.0104&	.0124&	.0074&	.0081&	.0095&	.0113\\
20000&	.0063&	.0070&	.0083&	.0100&	.0061&	.0067&	.0080&	.0095&	.0057&	.0062&	.0074&	.0088&	.0052&	.0057&	.0067&	.0080\\
50000&	.0040&	.0044&	.0053&	.0063&	.0038&	.0042&	.0050&	.0060&	.0036&	.0039&	.0047&	.0056&	.0033&	.0036&	.0043&	.0051\\
\noalign{\smallskip}											
\hline\noalign{\smallskip}																
& \multicolumn{4}{c|}{$K=100$: $\gamma=1.25$, Quantiles} & \multicolumn{4}{c|}{$\gamma=1.5$, Quantiles} & \multicolumn{4}{c|}{$\gamma=1.75$, Quantiles} & \multicolumn{4}{c}{$\gamma=2.0$, Quantiles} \\	
$N$ & 0.9 & 0.95 & 0.99 & 0.999 & 0.9 & 0.95 & 0.99 & 0.999 & 0.9 & 0.95 & 0.99 & 0.999 & 0.9 & 0.95 & 0.99 & 0.999 \\																
\noalign{\smallskip}\hline\noalign{\smallskip}																
10&	.2161&	.2381&	.2832&	.3405&	.2022&	.2259&	.2770&	.3420&	.1869&	.2115&	.2663&	.3366&	.1697&	.1941&	.2491&	.3212\\
20&	.1536&	.1691&	.2012&	.2424&	.1433&	.1599&	.1955&	.2443&	.1323&	.1496&	.1877&	.2349&	.1196&	.1370&	.1737&	.2232\\
30&	.1256&	.1384&	.1644&	.1977&	.1172&	.1310&	.1602&	.1991&	.1082&	.1224&	.1535&	.1931&	.0978&	.1116&	.1420&	.1803\\
40&	.1088&	.1199&	.1425&	.1723&	.1015&	.1134&	.1388&	.1727&	.0937&	.1059&	.1329&	.1684&	.0846&	.0966&	.1229&	.1573\\
50&	.0973&	.1072&	.1274&	.1543&	.0907&	.1014&	.1243&	.1548&	.0838&	.0948&	.1188&	.1511&	.0757&	.0864&	.1096&	.1407\\
100&	.0689&	.0758&	.0904&	.1090&	.0642&	.0717&	.0880&	.1096&	.0592&	.0670&	.0840&	.1062&	.0535&	.0610&	.0773&	.0982\\
500&	.0307&	.0339&	.0404&	.0487&	.0287&	.0320&	.0394&	.0493&	.0264&	.0299&	.0375&	.0473&	.0239&	.0273&	.0345&	.0434\\
1000&	.0217&	.0240&	.0286&	.0347&	.0203&	.0227&	.0279&	.0348&	.0187&	.0212&	.0266&	.0336&	.0169&	.0193&	.0244&	.0311\\
2000&	.0154&	.0169&	.0202&	.0244&	.0143&	.0160&	.0197&	.0246&	.0132&	.0150&	.0188&	.0237&	.0120&	.0137&	.0173&	.0219\\
3000&	.0126&	.0138&	.0165&	.0198&	.0117&	.0131&	.0161&	.0200&	.0108&	.0122&	.0153&	.0193&	.0098&	.0111&	.0141&	.0178\\
4000&	.0109&	.0120&	.0143&	.0173&	.0101&	.0113&	.0140&	.0175&	.0094&	.0106&	.0133&	.0168&	.0085&	.0096&	.0122&	.0155\\
5000&	.0097&	.0107&	.0128&	.0154&	.0091&	.0102&	.0125&	.0156&	.0084&	.0095&	.0119&	.0150&	.0076&	.0086&	.0109&	.0138\\
10000&	.0069&	.0076&	.0090&	.0110&	.0064&	.0072&	.0088&	.0109&	.0059&	.0067&	.0084&	.0106&	.0054&	.0061&	.0077&	.0098\\
20000&	.0049&	.0054&	.0064&	.0077&	.0045&	.0051&	.0062&	.0078&	.0042&	.0047&	.0059&	.0074&	.0038&	.0043&	.0055&	.0069\\
50000&	.0031&	.0034&	.0040&	.0049&	.0029&	.0032&	.0039&	.0049&	.0026&	.0030&	.0038&	.0047&	.0024&	.0027&	.0035&	.0044\\
\noalign{\smallskip}														
\hline\noalign{\smallskip}																
& \multicolumn{4}{c|}{$K=100$: $\gamma=2.5$, Quantiles} & \multicolumn{4}{c|}{$\gamma=3.0$, Quantiles} & \multicolumn{4}{c|}{$\gamma=3.5$, Quantiles} & \multicolumn{4}{c}{$\gamma=4.0$, Quantiles} \\	
$N$ & 0.9 & 0.95 & 0.99 & 0.999 & 0.9 & 0.95 & 0.99 & 0.999 & 0.9 & 0.95 & 0.99 & 0.999 & 0.9 & 0.95 & 0.99 & 0.999 \\																
\noalign{\smallskip}\hline\noalign{\smallskip}																
10&	.1335&	.1548&	.2093&	.2716&	.1001&	.1263&	.1655&	.2272&	.0819&	.0989&	.1356&	.1894&	.0819&	.0819&	.1088&	.1521\\
20&	.0942&	.1094&	.1410&	.1849&	.0721&	.0844&	.1111&	.1502&	.0534&	.0643&	.0878&	.1233&	.0440&	.0532&	.0734&	.0991\\
30&	.0764&	.0883&	.1140&	.1481&	.0571&	.0676&	.0892&	.1168&	.0429&	.0531&	.0698&	.0943&	.0311&	.0409&	.0550&	.0762\\
40&	.0660&	.0762&	.0981&	.1263&	.0498&	.0580&	.0762&	.0997&	.0372&	.0444&	.0592&	.0793&	.0281&	.0351&	.0463&	.0637\\
50&	.0590&	.0680&	.0874&	.1132&	.0445&	.0518&	.0674&	.0883&	.0330&	.0391&	.0525&	.0699&	.0264&	.0306&	.0408&	.0551\\
100&	.0417&	.0479&	.0612&	.0789&	.0314&	.0363&	.0467&	.0600&	.0234&	.0273&	.0356&	.0465&	.0178&	.0207&	.0276&	.0368\\
500&	.0186&	.0214&	.0273&	.0345&	.0140&	.0161&	.0205&	.0259&	.0105&	.0121&	.0154&	.0195&	.0079&	.0092&	.0118&	.0151\\
1000&	.0132&	.0152&	.0193&	.0245&	.0099&	.0114&	.0145&	.0183&	.0074&	.0085&	.0109&	.0137&	.0056&	.0064&	.0083&	.0105\\
2000&	.0093&	.0107&	.0136&	.0173&	.0070&	.0081&	.0103&	.0129&	.0052&	.0060&	.0077&	.0097&	.0039&	.0045&	.0058&	.0073\\
3000&	.0076&	.0087&	.0111&	.0141&	.0057&	.0066&	.0084&	.0105&	.0043&	.0049&	.0062&	.0079&	.0032&	.0037&	.0047&	.0059\\
4000&	.0066&	.0076&	.0096&	.0122&	.0049&	.0057&	.0072&	.0091&	.0037&	.0043&	.0054&	.0068&	.0028&	.0032&	.0041&	.0052\\
5000&	.0059&	.0068&	.0086&	.0109&	.0044&	.0051&	.0065&	.0082&	.0033&	.0038&	.0048&	.0061&	.0025&	.0029&	.0037&	.0046\\
10000&	.0042&	.0048&	.0061&	.0077&	.0031&	.0036&	.0046&	.0057&	.0023&	.0027&	.0034&	.0043&	.0018&	.0020&	.0026&	.0032\\
20000&	.0029&	.0034&	.0043&	.0055&	.0022&	.0026&	.0032&	.0041&	.0017&	.0019&	.0024&	.0030&	.0012&	.0014&	.0018&	.0023\\
50000&	.0019&	.0021&	.0027&	.0035&	.0014&	.0016&	.0020&	.0026&	.0010&	.0012&	.0015&	.0019&	.0008&	.0009&	.0012&	.0015\\
\noalign{\smallskip}\hline																
\end{tabular}																
\end{table*}		


\begin{table*}																
\caption{KS test statistic for the truncated power-law distribution with $K=500$ }			
\tiny																
\begin{tabular}{lrrrr|rrrr|rrrr|rrrr}
\hline\noalign{\smallskip}																
& \multicolumn{4}{c|}{$K=500$: $\gamma=0.25$, Quantiles} & \multicolumn{4}{c|}{$\gamma=0.5$, Quantiles} & \multicolumn{4}{c|}{$\gamma=0.75$, Quantiles} & \multicolumn{4}{c}{$\gamma=1.0$, Quantiles} \\
$N$ & 0.9 & 0.95 & 0.99 & 0.999 & 0.9 & 0.95 & 0.99 & 0.999 & 0.9 & 0.95 & 0.99 & 0.999 & 0.9 & 0.95 & 0.99 & 0.999 \\																
\noalign{\smallskip}\hline\noalign{\smallskip}																
10&	.2877&	.3160&	.3718&	.4371&	.2773&	.3048&	.3596&	.4237&	.2588&	.2837&	.3351&	.3972&	.2383&	.2599&	.3046&	.3595\\
20&	.2073&	.2279&	.2695&	.3183&	.1999&	.2195&	.2595&	.3068&	.1859&	.2038&	.2406&	.2841&	.1705&	.1860&	.2172&	.2556\\
30&	.1706&	.1876&	.2221&	.2632&	.1643&	.1807&	.2134&	.2533&	.1530&	.1676&	.1976&	.2345&	.1400&	.1527&	.1783&	.2101\\
40&	.1483&	.1632&	.1935&	.2307&	.1429&	.1571&	.1863&	.2216&	.1330&	.1458&	.1720&	.2042&	.1216&	.1326&	.1546&	.1823\\
50&	.1330&	.1465&	.1738&	.2061&	.1281&	.1410&	.1669&	.1983&	.1193&	.1307&	.1545&	.1831&	.1090&	.1188&	.1389&	.1631\\
100&	.0946&	.1041&	.1236&	.1468&	.0912&	.1003&	.1188&	.1412&	.0848&	.0930&	.1097&	.1301&	.0773&	.0844&	.0985&	.1157\\
500&	.0423&	.0466&	.0553&	.0656&	.0408&	.0449&	.0532&	.0632&	.0380&	.0417&	.0492&	.0583&	.0346&	.0378&	.0441&	.0518\\
1000&	.0299&	.0330&	.0390&	.0466&	.0289&	.0317&	.0376&	.0448&	.0269&	.0294&	.0347&	.0414&	.0245&	.0267&	.0311&	.0366\\
2000&	.0212&	.0233&	.0277&	.0329&	.0204&	.0224&	.0266&	.0317&	.0190&	.0208&	.0246&	.0293&	.0173&	.0189&	.0220&	.0259\\
3000&	.0173&	.0190&	.0225&	.0270&	.0167&	.0183&	.0217&	.0259&	.0155&	.0170&	.0201&	.0238&	.0141&	.0154&	.0180&	.0211\\
4000&	.0150&	.0165&	.0196&	.0234&	.0144&	.0159&	.0188&	.0225&	.0134&	.0147&	.0174&	.0207&	.0122&	.0134&	.0156&	.0183\\
5000&	.0134&	.0147&	.0175&	.0209&	.0129&	.0142&	.0168&	.0201&	.0120&	.0132&	.0155&	.0185&	.0110&	.0119&	.0139&	.0164\\
10000&	.0095&	.0104&	.0124&	.0147&	.0091&	.0101&	.0119&	.0142&	.0085&	.0093&	.0110&	.0130&	.0077&	.0084&	.0099&	.0116\\
20000&	.0067&	.0074&	.0088&	.0104&	.0065&	.0071&	.0084&	.0100&	.0060&	.0066&	.0078&	.0092&	.0055&	.0060&	.0070&	.0082\\
50000&	.0042&	.0047&	.0055&	.0066&	.0041&	.0045&	.0053&	.0063&	.0038&	.0042&	.0049&	.0058&	.0035&	.0038&	.0044&	.0052\\
\noalign{\smallskip}
\hline\noalign{\smallskip}																
& \multicolumn{4}{c|}{$K=500$: $\gamma=1.25$, Quantiles} & \multicolumn{4}{c|}{$\gamma=1.5$, Quantiles} & \multicolumn{4}{c|}{$\gamma=1.75$, Quantiles} & \multicolumn{4}{c}{$\gamma=2.0$, Quantiles} \\	
$N$ & 0.9 & 0.95 & 0.99 & 0.999 & 0.9 & 0.95 & 0.99 & 0.999 & 0.9 & 0.95 & 0.99 & 0.999 & 0.9 & 0.95 & 0.99 & 0.999 \\																
\noalign{\smallskip}\hline\noalign{\smallskip}																
10&	.2259&	.2487&	.2961&	.3548&	.2150&	.2409&	.2954&	.3637&	.1987&	.2251&	.2834&	.3560&	.1780&	.2055&	.2621&	.3370\\
20&	.1609&	.1772&	.2108&	.2533&	.1525&	.1708&	.2101&	.2591&	.1405&	.1595&	.2002&	.2510&	.1255&	.1438&	.1832&	.2343\\
30&	.1317&	.1450&	.1728&	.2069&	.1248&	.1400&	.1722&	.2118&	.1149&	.1304&	.1637&	.2045&	.1025&	.1173&	.1493&	.1899\\
40&	.1142&	.1257&	.1500&	.1808&	.1081&	.1212&	.1490&	.1853&	.0996&	.1129&	.1418&	.1798&	.0887&	.1015&	.1291&	.1646\\
50&	.1022&	.1125&	.1339&	.1618&	.0968&	.1084&	.1336&	.1658&	.0891&	.1011&	.1268&	.1597&	.0794&	.0908&	.1153&	.1481\\
100&	.0723&	.0797&	.0953&	.1151&	.0685&	.0767&	.0947&	.1182&	.0629&	.0715&	.0897&	.1130&	.0561&	.0640&	.0813&	.1032\\
500&	.0324&	.0357&	.0426&	.0518&	.0306&	.0343&	.0422&	.0528&	.0281&	.0319&	.0401&	.0505&	.0251&	.0287&	.0363&	.0456\\
1000&	.0229&	.0252&	.0301&	.0366&	.0217&	.0243&	.0299&	.0374&	.0199&	.0226&	.0283&	.0357&	.0177&	.0203&	.0257&	.0325\\
2000&	.0162&	.0178&	.0213&	.0258&	.0153&	.0172&	.0212&	.0264&	.0141&	.0160&	.0200&	.0252&	.0125&	.0143&	.0182&	.0230\\
3000&	.0132&	.0146&	.0174&	.0210&	.0125&	.0140&	.0173&	.0214&	.0115&	.0131&	.0164&	.0205&	.0102&	.0117&	.0148&	.0187\\
4000&	.0114&	.0126&	.0151&	.0183&	.0108&	.0121&	.0150&	.0188&	.0100&	.0113&	.0142&	.0179&	.0089&	.0101&	.0128&	.0162\\
5000&	.0102&	.0113&	.0135&	.0163&	.0097&	.0108&	.0134&	.0167&	.0089&	.0101&	.0127&	.0160&	.0079&	.0091&	.0115&	.0145\\
10000&	.0072&	.0080&	.0095&	.0115&	.0068&	.0077&	.0095&	.0117&	.0063&	.0071&	.0090&	.0113&	.0056&	.0064&	.0081&	.0103\\
20000&	.0051&	.0056&	.0067&	.0082&	.0048&	.0054&	.0067&	.0083&	.0045&	.0051&	.0063&	.0079&	.0040&	.0045&	.0057&	.0072\\
50000&	.0032&	.0036&	.0043&	.0052&	.0031&	.0034&	.0042&	.0052&	.0028&	.0032&	.0040&	.0050&	.0025&	.0029&	.0036&	.0046\\
\noalign{\smallskip}
\hline\noalign{\smallskip}																
& \multicolumn{4}{c|}{$K=500$: $\gamma=2.5$, Quantiles} & \multicolumn{4}{c|}{$\gamma=3.0$, Quantiles} & \multicolumn{4}{c|}{$\gamma=3.5$, Quantiles} & \multicolumn{4}{c}{$\gamma=4.0$, Quantiles} \\	
$N$ & 0.9 & 0.95 & 0.99 & 0.999 & 0.9 & 0.95 & 0.99 & 0.999 & 0.9 & 0.95 & 0.99 & 0.999 & 0.9 & 0.95 & 0.99 & 0.999 \\																
\noalign{\smallskip}\hline\noalign{\smallskip}																
10&	.1362&	.1584&	.2123&	.2809&	.1004&	.1278&	.1688&	.2332&	.0819&	.0992&	.1363&	.1915&	.0819&	.0819&	.1091&	.1529\\
20&	.0959&	.1115&	.1457&	.1914&	.0723&	.0848&	.1122&	.1525&	.0534&	.0642&	.0879&	.1245&	.0440&	.0533&	.0734&	.0994\\
30&	.0779&	.0898&	.1168&	.1522&	.0573&	.0678&	.0897&	.1183&	.0430&	.0532&	.0699&	.0949&	.0311&	.0409&	.0550&	.0764\\
40&	.0672&	.0778&	.1005&	.1301&	.0501&	.0584&	.0770&	.1011&	.0372&	.0444&	.0593&	.0799&	.0281&	.0351&	.0463&	.0639\\
50&	.0601&	.0694&	.0895&	.1161&	.0447&	.0520&	.0678&	.0897&	.0330&	.0391&	.0526&	.0703&	.0264&	.0306&	.0408&	.0552\\
100&	.0425&	.0489&	.0627&	.0808&	.0315&	.0365&	.0471&	.0607&	.0235&	.0273&	.0357&	.0467&	.0178&	.0207&	.0277&	.0369\\
500&	.0190&	.0218&	.0278&	.0351&	.0141&	.0162&	.0207&	.0261&	.0105&	.0121&	.0155&	.0196&	.0079&	.0092&	.0118&	.0151\\
1000&	.0134&	.0154&	.0197&	.0250&	.0100&	.0115&	.0145&	.0185&	.0074&	.0086&	.0109&	.0137&	.0056&	.0064&	.0083&	.0105\\
2000&	.0095&	.0109&	.0139&	.0177&	.0070&	.0081&	.0103&	.0130&	.0052&	.0060&	.0077&	.0097&	.0039&	.0046&	.0058&	.0073\\
3000&	.0077&	.0089&	.0113&	.0144&	.0057&	.0066&	.0084&	.0106&	.0043&	.0049&	.0063&	.0079&	.0032&	.0037&	.0047&	.0059\\
4000&	.0067&	.0077&	.0098&	.0124&	.0050&	.0057&	.0073&	.0092&	.0037&	.0043&	.0054&	.0068&	.0028&	.0032&	.0041&	.0052\\
5000&	.0060&	.0069&	.0088&	.0112&	.0045&	.0051&	.0065&	.0082&	.0033&	.0038&	.0048&	.0061&	.0025&	.0029&	.0037&	.0046\\
10000&	.0042&	.0049&	.0062&	.0079&	.0031&	.0036&	.0046&	.0058&	.0023&	.0027&	.0034&	.0043&	.0018&	.0020&	.0026&	.0032\\
20000&	.0030&	.0034&	.0044&	.0056&	.0022&	.0026&	.0033&	.0041&	.0017&	.0019&	.0024&	.0030&	.0012&	.0014&	.0018&	.0023\\
50000&	.0019&	.0022&	.0028&	.0035&	.0014&	.0016&	.0021&	.0026&	.0010&	.0012&	.0015&	.0019&	.0008&	.0009&	.0012&	.0015\\
\noalign{\smallskip}\hline																
\end{tabular}																
\end{table*}


\begin{table*}																
\caption{KS test statistic for the truncated power-law distribution with $K=1000$ }\label{table-last}
\tiny																
\begin{tabular}{lrrrr|rrrr|rrrr|rrrr}
\hline\noalign{\smallskip}																
& \multicolumn{4}{c|}{$K=1000$: $\gamma=0.25$, Quantiles} & \multicolumn{4}{c|}{$\gamma=0.5$, Quantiles} & \multicolumn{4}{c|}{$\gamma=0.75$, Quantiles} &\multicolumn{4}{c}{$\gamma=1.0$, Quantiles}\\
$N$ & 0.9 & 0.95 & 0.99 & 0.999 & 0.9 & 0.95 & 0.99 & 0.999 & 0.9 & 0.95 & 0.99 & 0.999 & 0.9 & 0.95 & 0.99 & 0.999 \\																
\noalign{\smallskip}\hline\noalign{\smallskip}																
10&	.2901&	.3184&	.3745&	.4403&	.2806&	.3082&	.3635&	.4279&	.2616&	.2868&	.3387&	.4000&	.2402&	.2617&	.3062&	.3604\\
20&	.2091&	.2299&	.2717&	.3207&	.2023&	.2223&	.2627&	.3103&	.1881&	.2063&	.2435&	.2873&	.1719&	.1874&	.2188&	.2568\\
30&	.1722&	.1894&	.2239&	.2656&	.1664&	.1830&	.2164&	.2565&	.1549&	.1697&	.2001&	.2375&	.1413&	.1539&	.1795&	.2110\\
40&	.1497&	.1647&	.1952&	.2322&	.1448&	.1592&	.1887&	.2244&	.1347&	.1476&	.1743&	.2073&	.1227&	.1337&	.1557&	.1841\\
50&	.1343&	.1479&	.1754&	.2077&	.1298&	.1429&	.1691&	.2009&	.1208&	.1325&	.1565&	.1851&	.1100&	.1198&	.1398&	.1645\\
100&	.0957&	.1053&	.1248&	.1484&	.0925&	.1017&	.1204&	.1429&	.0860&	.0943&	.1112&	.1319&	.0781&	.0851&	.0992&	.1164\\
500&	.0430&	.0472&	.0561&	.0666&	.0416&	.0457&	.0541&	.0642&	.0387&	.0424&	.0500&	.0594&	.0350&	.0382&	.0445&	.0522\\
1000&	.0304&	.0334&	.0395&	.0470&	.0294&	.0323&	.0382&	.0455&	.0273&	.0299&	.0353&	.0420&	.0248&	.0270&	.0315&	.0369\\
2000&	.0215&	.0236&	.0280&	.0333&	.0208&	.0228&	.0271&	.0322&	.0193&	.0212&	.0250&	.0297&	.0175&	.0191&	.0222&	.0260\\
3000&	.0175&	.0193&	.0228&	.0273&	.0170&	.0186&	.0221&	.0263&	.0158&	.0173&	.0204&	.0242&	.0143&	.0156&	.0182&	.0213\\
4000&	.0152&	.0167&	.0198&	.0237&	.0147&	.0161&	.0191&	.0228&	.0137&	.0150&	.0177&	.0210&	.0124&	.0135&	.0158&	.0185\\
5000&	.0136&	.0149&	.0177&	.0211&	.0131&	.0144&	.0171&	.0204&	.0122&	.0134&	.0158&	.0188&	.0111&	.0121&	.0141&	.0165\\
10000&	.0096&	.0106&	.0125&	.0149&	.0093&	.0102&	.0121&	.0144&	.0086&	.0095&	.0112&	.0132&	.0078&	.0085&	.0100&	.0117\\
20000&	.0068&	.0075&	.0089&	.0105&	.0066&	.0072&	.0086&	.0102&	.0061&	.0067&	.0079&	.0094&	.0055&	.0060&	.0070&	.0083\\
50000&	.0043&	.0047&	.0056&	.0067&	.0042&	.0046&	.0054&	.0065&	.0039&	.0042&	.0050&	.0059&	.0035&	.0038&	.0044&	.0052\\
\noalign{\smallskip}
\hline\noalign{\smallskip}																
& \multicolumn{4}{c|}{$K=1000$: $\gamma=1.25$, Quantiles} & \multicolumn{4}{c|}{$\gamma=1.5$, Quantiles} & \multicolumn{4}{c|}{$\gamma=1.75$, Quantiles} & \multicolumn{4}{c}{$\gamma=2.0$, Quantiles} \\
$N$ & 0.9 & 0.95 & 0.99 & 0.999 & 0.9 & 0.95 & 0.99 & 0.999 & 0.9 & 0.95 & 0.99 & 0.999 & 0.9 & 0.95 & 0.99 & 0.999 \\																
\noalign{\smallskip}\hline\noalign{\smallskip}																
10&	.2296&	.2529&	.3016&	.3594&	.2197&	.2464&	.3019&	.3696&	.2022&	.2295&	.2884&	.3610&	.1800&	.2077&	.2662&	.3408\\
20&	.1635&	.1801&	.2145&	.2581&	.1558&	.1747&	.2144&	.2643&	.1427&	.1622&	.2036&	.2551&	.1268&	.1455&	.1854&	.2369\\
30&	.1338&	.1475&	.1759&	.2119&	.1275&	.1430&	.1759&	.2162&	.1168&	.1325&	.1668&	.2079&	.1036&	.1185&	.1510&	.1914\\
40&	.1161&	.1279&	.1525&	.1841&	.1105&	.1239&	.1522&	.1894&	.1012&	.1148&	.1444&	.1824&	.0895&	.1026&	.1306&	.1664\\
50&	.1039&	.1144&	.1363&	.1646&	.0988&	.1108&	.1366&	.1683&	.0906&	.1028&	.1290&	.1624&	.0802&	.0917&	.1166&	.1496\\
100&	.0736&	.0811&	.0970&	.1173&	.0699&	.0784&	.0968&	.1209&	.0640&	.0726&	.0913&	.1146&	.0566&	.0647&	.0821&	.1045\\
500&	.0329&	.0363&	.0434&	.0525&	.0313&	.0351&	.0432&	.0541&	.0286&	.0324&	.0407&	.0511&	.0253&	.0290&	.0367&	.0461\\
1000&	.0233&	.0257&	.0307&	.0372&	.0221&	.0248&	.0306&	.0382&	.0202&	.0230&	.0288&	.0363&	.0179&	.0205&	.0259&	.0328\\
2000&	.0165&	.0182&	.0217&	.0262&	.0156&	.0176&	.0217&	.0269&	.0143&	.0162&	.0204&	.0256&	.0127&	.0145&	.0184&	.0232\\
3000&	.0134&	.0148&	.0177&	.0214&	.0128&	.0143&	.0176&	.0219&	.0117&	.0133&	.0166&	.0208&	.0103&	.0118&	.0150&	.0189\\
4000&	.0116&	.0128&	.0153&	.0186&	.0110&	.0124&	.0153&	.0191&	.0101&	.0115&	.0144&	.0181&	.0089&	.0102&	.0129&	.0164\\
5000&	.0104&	.0115&	.0137&	.0166&	.0099&	.0111&	.0137&	.0170&	.0090&	.0103&	.0129&	.0162&	.0080&	.0092&	.0116&	.0147\\
10000&	.0074&	.0081&	.0097&	.0117&	.0070&	.0078&	.0097&	.0120&	.0064&	.0073&	.0091&	.0115&	.0057&	.0065&	.0082&	.0104\\
20000&	.0052&	.0057&	.0069&	.0083&	.0049&	.0056&	.0068&	.0085&	.0045&	.0051&	.0064&	.0081&	.0040&	.0046&	.0058&	.0073\\
50000&	.0033&	.0036&	.0043&	.0052&	.0031&	.0035&	.0043&	.0054&	.0029&	.0032&	.0041&	.0051&	.0025&	.0029&	.0037&	.0046\\
\noalign{\smallskip}
\hline\noalign{\smallskip}																
& \multicolumn{4}{c|}{$K=1000$: $\gamma=2.5$, Quantiles} & \multicolumn{4}{c|}{$\gamma=3.0$, Quantiles} & \multicolumn{4}{c|}{$\gamma=3.5$, Quantiles} & \multicolumn{4}{c}{$\gamma=4.0$, Quantiles} \\
$N$ & 0.9 & 0.95 & 0.99 & 0.999 & 0.9 & 0.95 & 0.99 & 0.999 & 0.9 & 0.95 & 0.99 & 0.999 & 0.9 & 0.95 & 0.99 & 0.999 \\	
\noalign{\smallskip}\hline\noalign{\smallskip}
10&	.1368&	.1587&	.2131&	.2843&	.1005&	.1279&	.1692&	.2340&	.0819&	.0992&	.1364&	.1918&	.0819&	.0819&	.1092&	.1530\\
20&	.0961&	.1116&	.1464&	.1924&	.0723&	.0849&	.1123&	.1527&	.0534&	.0642&	.0879&	.1245&	.0440&	.0533&	.0734&	.0994\\
30&	.0780&	.0900&	.1172&	.1531&	.0573&	.0678&	.0897&	.1185&	.0430&	.0532&	.0699&	.0950&	.0311&	.0409&	.0550&	.0763\\
40&	.0673&	.0780&	.1009&	.1307&	.0501&	.0584&	.0770&	.1013&	.0372&	.0444&	.0593&	.0800&	.0281&	.0351&	.0463&	.0639\\
50&	.0602&	.0696&	.0898&	.1167&	.0447&	.0520&	.0679&	.0897&	.0330&	.0391&	.0526&	.0703&	.0264&	.0306&	.0408&	.0552\\
100&	.0426&	.0490&	.0629&	.0812&	.0315&	.0365&	.0471&	.0608&	.0235&	.0273&	.0357&	.0467&	.0178&	.0207&	.0277&	.0369\\
500&	.0190&	.0218&	.0279&	.0352&	.0141&	.0162&	.0207&	.0261&	.0105&	.0121&	.0155&	.0196&	.0079&	.0092&	.0118&	.0151\\
1000&	.0134&	.0155&	.0198&	.0251&	.0100&	.0115&	.0146&	.0185&	.0074&	.0086&	.0109&	.0137&	.0056&	.0064&	.0083&	.0105\\
2000&	.0095&	.0109&	.0139&	.0177&	.0070&	.0081&	.0103&	.0130&	.0052&	.0060&	.0077&	.0097&	.0039&	.0046&	.0058&	.0073\\
3000&	.0077&	.0089&	.0114&	.0144&	.0057&	.0066&	.0084&	.0106&	.0043&	.0049&	.0063&	.0079&	.0032&	.0037&	.0047&	.0059\\
4000&	.0067&	.0077&	.0098&	.0125&	.0050&	.0057&	.0073&	.0092&	.0037&	.0043&	.0054&	.0068&	.0028&	.0032&	.0041&	.0052\\
5000&	.0060&	.0069&	.0088&	.0112&	.0045&	.0051&	.0065&	.0082&	.0033&	.0038&	.0048&	.0061&	.0025&	.0029&	.0037&	.0046\\
10000&	.0042&	.0049&	.0062&	.0079&	.0031&	.0036&	.0046&	.0058&	.0023&	.0027&	.0034&	.0043&	.0018&	.0020&	.0026&	.0032\\
20000&	.0030&	.0034&	.0044&	.0056&	.0022&	.0026&	.0033&	.0041&	.0017&	.0019&	.0024&	.0030&	.0012&	.0014&	.0018&	.0023\\
50000&	.0019&	.0022&	.0028&	.0035&	.0014&	.0016&	.0021&	.0026&	.0010&	.0012&	.0015&	.0019&	.0008&	.0009&	.0012&	.0015\\
\noalign{\smallskip}\hline
\end{tabular}
\end{table*}

\clearpage

\lstset{
language=C++,
basicstyle=\scriptsize\sffamily,
numbers=left,
numberstyle=\tiny,
frame=tb,
columns=fullflexible,
showstringspaces=false,
tabsize=3
}

\begin{lstlisting}[label=label, caption=CUDA/C code]

#include <stdio.h>
#include <stdlib.h>
#include <time.h>
#include "cuda_runtime.h"
#include "device_launch_parameters.h"
#include <cuda_runtime_api.h>
#include <curand.h>
#include <curand_kernel.h>

/* 
	--------------------------------------------------------------------------------

	CUDA C program used for the results of the article 
	
	DISCRETE TRUNCATED ZIPF DISTRIBUTION:

	Calculates the quantiles for a given value of K, gamma, and random seed. 
	
	Syntax is: 
	program.exe  K  Gamma  Random_Seed_integer

	The value of N is fixed in the code.

	- K must be less than 32766 (in the paper, it's 20, 30, 50, 100, 500, 1000).
	- The value of N is fixed at the start of the code below. 
	  In the paper, it's 10, 20, 30, 40, 50, 100, 500, 1000, 2000, 3000, 4000, 5000, 10000, 20000. 
	- Gamma should be >0.25 (for meaningful results)

	----------------------------------------------------------------------------------

	Technical note: Important when compiling this CUDA program:
	
	- The program requires a GPU that supports CUDA, and the (freely downloadable) CUDA developer software installed 
	  (for Visual Studio it is available as an add-on)
	- The program requires a CUDA GPU that supports 'double' floating-point numbers. Some GPU only support 'float' - this is not good enough 
	  and will produce incorrect results (zeros, infinities) due to the accumulation of rounding errors
	- By default, CUDA may demote 'double' to 'float' to conserve resources. If a compilation warning:
	  'double is not supported, demoting to float' is produced, the following compilation parameters need to be adjusted:
	  code generation  = compute_20, sm_20
	  compiler options = -arch=sm_20
	  (20 refers to the cude computational ability level of the card; level 13 or more supports 'double')
	- The standard CUDA libaray curand.lib must be included (used for random number generation)

	-----------------------------------------------------------------------------------

	Dr. Efstratios Rappos and Prof. Stephan Robert
	HEIG-VD 
	Switzerland

	Efstratios.Rappos at heig-vd <dot> ch
	Stephan.Robert at heig-vd <dot> ch

	May 2013  

*/


// The number of points in each simulation (sample size N)

#define N 2000

#define CUDA_GPU_DEVICE 2  // If you have multiple NVIDIA cards, specify which to use. Start with 0 = "first card", 1 = "second card" etc.

#define BLOCKS				782  
#define THREADS_PER_BLOCK   64  

#define SIMULATIONS  BLOCKS * THREADS_PER_BLOCK   //number of simulations  (a multiple of NTHREADS)
#define SIMULATIONS_REQUIRED 50000

/* 
	SIM =   1 *  64 =    64
	SIM =   2 *  64 =   128
	SIM =   4 *  64 =   256
	SIM =   5 *  64 =   320
	SIM =   7 *  64 =   448
	SIM =   8 *  64 =   512
	SIM =  16 *  64 =  1024
	SIM =  32 *  64 =  2048
	SIM =  63 *  64 =  4032
	SIM =  79 *  64 =  5056
	SIM = 157 *  64 = 10048
	SIM = 313 *  64 = 20032
	SIM = 782 *  64 = 50048
*/


// Nothing really to change from here on.

//#define MAX_POINTS 32767 

int sort_dbl(const void *x, const void *y) {
		double t = (*(double*)x - *(double*)y);
		return (int) ( (t>0) - (t<0)) ;
 }

__device__  double NewtonRaphson(double initial_guess, double RHS_data,  int K, const double * LOGS);
__device__  double KolmogorovSmirnoff_short(const unsigned short * data,   int K, double gamma, const double * LOGS);

__host__ void check_cuda(cudaError_t cudaStatus, char* message, bool &fail){

		if(cudaStatus){ 
			printf("Error in %s  (%d) - %s\n ", message, cudaStatus, cudaGetErrorString(cudaStatus)); 
			fail = true;
			system("pause");
		}
}


__global__ void setup_kernel (int seed, curandState * state ){

	int id = threadIdx .x + blockIdx .x *  THREADS_PER_BLOCK ;
	unsigned long long seed1 = seed;
	// Each thread gets the same seed, but a different sequence number, no offset 
	curand_init (seed1 , id , 0, &state[id]);
}

__global__ 	void generate_kernel ( curandState *state ,  double * dev_results,  const int K, const double gamma,  const double *LOGS){

	int id = threadIdx .x + blockIdx .x * THREADS_PER_BLOCK;
	curandState localState = state[id];    // Copy state to local memory for efficiency 
	unsigned short points[N];
	int i, t;
	double x, c;

	for(i=0;i<N;i++)
		points[i] = 0;

	int KMAX=0;
	c = 0.0;
	
	for (i=1; i<=K; i++)
		c = c + exp( - (double) gamma * LOGS[i]);  // c = c + (1.0 / pow((double) i, (double) gamma));
	c = 1.0 / c;

	for(t=0; t<N; t++) {		
		x = curand_uniform_double (& localState );
		double sum_prob = 0;
		for (i=1;  ; i++){
			sum_prob = sum_prob + c*exp(-(double)gamma * LOGS[i] );
			if (sum_prob >= x){
				points[t]= i;  
				if(i>KMAX) KMAX = i;
				break;
			}
		}
	}

	// We store the value of KMAX, the max observation in the current generated series. 
	// As all observations are <=K anyway (an input parameter), we will have KMAX <= K, 
	// However, when using loops 1 to K, we can loop up to KMAX only, rather than K, as there are no observations in the range KMAX to K (more efficient).

	// Copy state back to global memory 
	state[id] = localState ;

	// We now have points[]

	// FIRST FIND Maximum Likelihood Estimator
	// RHS

	int NPOINTS=0;
	double RHS_data = 0.0;
	
	for(t=0;t<N;t++){
		// check: should never happen
		if((points[t]<1)||(points[t]>32766)) {printf("\n\n ERROR  points[%d] < 1 (=%d) !\n\n",t,points[t]); dev_results[id] = -1.; return; }
			
		RHS_data +=  LOGS[points[t]] ; 
		NPOINTS++;
	}


// If all points are = 1, adjust RHS so that it's >0..(for RHS=0 the estimated gamma is infinity). 
// adjust by a factor of ln(2) -- as if one point was 2 instead of 1. This given a max gamma of ~15 for 50,000 total points
// This only affects very small values of N, eg 10, 20, 30, 40.

//	if(RHS_data<=0){ printf("RHS is <=0 ! (%f)\nPOINTS=",RHS_data);
//		for(t=0;t<N;t++) printf("%d,",points[t]); printf(".-\n");
//	}

	if(RHS_data<=0) 
		RHS_data += LOGS[2];
	
	RHS_data = RHS_data / (double) NPOINTS;

	// Newton-Raphson to obtain estimated value for gamma
	double estimated_gamma  =  NewtonRaphson( 0.5 ,  RHS_data ,  K, LOGS); //must be K, not KMAX
	
	//Kolmogorov-Smirnoff test statistic
	double KStest = KolmogorovSmirnoff_short(points, K, estimated_gamma, LOGS);
	dev_results[id] = KStest;
}

int main(int argc, char* argv[]){

	if(argc != 4){
		printf("syntax is program.exe K GAMMA SEED\nbye\n");
		system("pause");
		return 1;
	}

	const int K = atoi(argv[1]); 
	if(K==0){
		printf("Cannot read value for K\nbye\n");
		system("pause");
		return 1;
	}

	const double gamma = atof(argv[2]);
	if(gamma<0.1){
		printf("Cannot read value for Gamma, or Gamma<0.1\nbye\n");
		system("pause");
		return 1;
	}

	const int seed = atoi(argv[3]);
	if(seed<1){
		printf("Cannot read value for SEED, or SEED<1\nbye\n");
		system("pause");
		return 1;
	}

	if(K>32766){
		printf("Value for K must be < 32766\nbye\n"); //must be < 32,767  as with the fast implementation, the CUDA sample points are coded 'short'
		system("pause");
		return 1;
	}

	if(SIMULATIONS_REQUIRED > SIMULATIONS){
		printf("SIMULATIONS_REQUIRED must be <= SIMULATIONS\nbye\n"); 
		system("pause");
		return 1;
	}

	cudaError_t cudaStatus ;

	int i;

	// Pre-compute Logarithms for 1--K, K < 32767, for faster execution

	double * LOGS; 

	LOGS = new double[K+2];
	for(i=0;i<=(K+1);i++)
		LOGS[i] = log((double) i);
	
	cudaGetDeviceCount(&i);
	printf("Found %d Graphics cards that support CUDA\n",i);
	printf("Checking capabilities of chosen GPU device (CUDA_GPU_DEVICE = %d):\n",CUDA_GPU_DEVICE);
	
	cudaDeviceProp properties;
	cudaGetDeviceProperties(&properties, CUDA_GPU_DEVICE);
		printf("   Name:                             %s\n", properties.name);
		printf("   Total global Memory:              %d\n", (int) properties.totalGlobalMem);
		printf("   Shared Memory per block:          %d\n", (int) properties.sharedMemPerBlock);
		printf("   Total Const Memory:               %d\n", (int) properties.totalConstMem);
		printf("   Multiprocessors:                  %d\n", properties.multiProcessorCount);
		printf("   Max # threads per multiprocessor: %d\n", properties.maxThreadsPerMultiProcessor);
		printf("   Max #threads per per block:       %d\n", (int) properties.maxThreadsPerBlock);
		printf("   Compute capability:               %d.%d\n", properties.major, properties.minor);
		printf("   Kernel timeout enabled:           %d\n", properties.kernelExecTimeoutEnabled);

	bool fail = false;
		
	cudaStatus = cudaSetDevice(CUDA_GPU_DEVICE);	check_cuda(cudaStatus, "setdevice", fail);

	time_t t1 = clock();

	// Copy logarithms to CUDA device
	double *dev_logs = 0;
	cudaStatus = cudaMalloc((void**)&dev_logs, (K+2) * sizeof(double));		
	cudaStatus = cudaMemcpy(dev_logs, LOGS, (K+2) * sizeof(double), cudaMemcpyHostToDevice);
	
	delete[] LOGS;

	// Simulation Setup	
	double KStest_sim[SIMULATIONS];  //stores the K-Smirnoff statisitc
	for(i=0;i<SIMULATIONS;i++)
		KStest_sim[i] = -1.0 ;
   
	printf("Generating %d power-law distributions, with N=%d, K=%d, gamma=%f \n", SIMULATIONS, N, K, gamma );
	// generate #SIMULATIONS random seed values using the CUDA random generator
	curandState * devStates ;
	cudaMalloc((void **)&devStates, SIMULATIONS * sizeof(curandState));

			setup_kernel <<<BLOCKS, THREADS_PER_BLOCK>>>( seed, devStates );   

	cudaStatus = cudaDeviceSynchronize();														 check_cuda(cudaStatus, "cudaMemcpy 2", fail); 
	cudaStatus = cudaGetLastError();															 check_cuda(cudaStatus, "lastError i", fail); 

	double * dev_results;
	cudaMalloc((void**)&dev_results, SIMULATIONS * sizeof(double));			check_cuda(cudaStatus, "cudaMemcpy 4", fail); 
	cudaMemset(dev_results, 0, SIMULATIONS * sizeof(double));				check_cuda(cudaStatus, "cudaMemcpy 5", fail);
	cudaStatus =	cudaGetLastError();										check_cuda(cudaStatus, "lastError ii", fail);

			generate_kernel <<<BLOCKS, THREADS_PER_BLOCK>>>( devStates ,  dev_results, K,  gamma  , dev_logs );  

	cudaStatus =	cudaGetLastError();											 check_cuda(cudaStatus, "lastError iii", fail); 
	cudaStatus = cudaDeviceSynchronize();										 check_cuda(cudaStatus, "cudaDeviceSunchronize 6", fail); 
	cudaStatus =	cudaGetLastError();											 check_cuda(cudaStatus, "lastError iv 2", fail); 


	cudaMemcpy (KStest_sim, dev_results, SIMULATIONS * sizeof(double), cudaMemcpyDeviceToHost);	 check_cuda(cudaStatus, "cudaMemcpy 7", fail); 	
	
	cudaFree(dev_logs);
	cudaFree(dev_results);
	cudaFree(devStates);


	// error catching - should never happen	
	//for(i=0;i<SIMULATIONS;i++){
	//	if(KStest_sim[i]<0.000){
	//		printf("\n\n KStest_sim is <0.0  (%f)!!\n", KStest_sim[i]);
	//		return 1;
	//	}
	//}
	

	//Calculate Quantiles

	// As # simulations is a multiple of 64, we must discard some sumulaitons to have the required number

	double KStest2[SIMULATIONS_REQUIRED];

	for(i=0;i<SIMULATIONS_REQUIRED;i++)
		KStest2[i] = KStest_sim[i]; 
	
	qsort(KStest2, SIMULATIONS_REQUIRED, sizeof(double), sort_dbl);
				
	printf("Quantile 90  %% is at %6.4f \n", KStest2[ SIMULATIONS_REQUIRED*9/10     ]);
	printf("Quantile 95  %% is at %6.4f \n", KStest2[ SIMULATIONS_REQUIRED*95/100   ]);
	printf("Quantile 99  %% is at %6.4f \n", KStest2[ SIMULATIONS_REQUIRED*99/100   ]);
	printf("Quantile 99.9%% is at %6.4f \n", KStest2[ SIMULATIONS_REQUIRED*999/1000 ]);
		
	time_t t2 = clock();
	double duration = (double)(t2-t1) / CLOCKS_PER_SEC;

	// if output to a text file is desired
	FILE *fout;
	fout = fopen("output.txt", "a+");

	fprintf(fout, "%d & %d & %6.2f & %6.4f & %6.4f & %6.4f & %6.4f & %10.4f\\\\\n",K, N, gamma, 
		KStest2[ SIMULATIONS_REQUIRED*9/10     ], 
		KStest2[ SIMULATIONS_REQUIRED*95/100   ], 
		KStest2[ SIMULATIONS_REQUIRED*99/100   ], 
		KStest2[ SIMULATIONS_REQUIRED*999/1000 ],
		duration);
	
	fclose(fout);	
	printf("Time taken:  %10.4f seconds\n", duration);
	return 0;
}



// Newton - Rapshson algorithm: produces the estimate the power-law exponent gamma from the data 

__device__  double NewtonRaphson(double initial_guess, double RHS_data, int K, const double * LOGS){

		const double absolute_tolerance = 0.00001;  // the required level of accuracy in the estimation of gamma
		int t;
		double x, xnew;
		x = xnew = initial_guess;    		// initial guesses for gamma
		double A, B, C;		 
				
		do{

			x = xnew;
			double f, f1;
			f = 0.0;

			A=0.0; B=0.0; C=0.0;
			
			for(t=1;t<=K;t++){

				double powt = exp(- x * LOGS[t]);

				A += (-  powt * LOGS[t]  );    
				B +=  powt;  // C
				C +=  powt * LOGS[t]  * LOGS[t] ; 			
			}

			//f(x)
			f = A/B + RHS_data ;

			//f'(x) - the derivative
			f1 = C/B - A/B*A/B;
			xnew = x - f / f1;
		}

		while(( abs(x - xnew) > absolute_tolerance));  	
	
		return xnew;
}



// Kolmogorov - Smirnoff test:  returns the test value of the test
__device__  double KolmogorovSmirnoff_short(const unsigned short * data, int K, double gamma, const double * LOGS){


		double c = 0.0;
		int i; 
		
		int t;
		double xnew = gamma;


		for(t=1;t<=K;t++)		
			c +=   exp( -xnew *LOGS[t]); 

		c = 1.0 / c;
		
		double actual_prev, theoretical_prev;
		int actual;
		double KStest = -2.0;

		int NPOINTS = N;
		
		for(t=1;t<=K;t++){  //K here is the max observation

			if(t==1){
				
				theoretical_prev = c *  exp( -xnew * LOGS[t]) ;     
				actual = 0;
				for(i=0;i<NPOINTS;i++){
					if(data[i] == t)
						actual++;
				}
				actual_prev = (double) actual / (double) NPOINTS;
			}
			else {
				
				theoretical_prev += c *   exp( -xnew * LOGS[t]) ;  
				actual = 0;
				for(i=0;i<NPOINTS;i++){
					if(data[i] == t)
						actual++;
				}
				actual_prev += (double) actual / (double) NPOINTS;
			}
	
			// Find SUP
			if(abs(theoretical_prev - actual_prev) >KStest  )
				KStest = abs(theoretical_prev - actual_prev);
		}
		return KStest;
}

\end{lstlisting}

\end{document}